\documentclass[aps,prl,reprint,groupedaddress]{revtex4-1}

\usepackage{graphicx}
\usepackage{dcolumn}

\begin{document}


\title{Simultaneous Dual-Species Matter-Wave Accelerometer}


\author{A. Bonnin}
\email{alexis.bonnin@onera.fr}
\affiliation{ONERA, DMPH, BP 80100, 91123, Palaiseau, France}

\author{N. Zahzam}
\affiliation{ONERA, DMPH, BP 80100, 91123, Palaiseau, France}

\author{Y. Bidel}
\affiliation{ONERA, DMPH, BP 80100, 91123, Palaiseau, France}

\author{A. Bresson}
\affiliation{ONERA, DMPH, BP 80100, 91123, Palaiseau, France}


\date{\today}

\begin{abstract}
We report the realization of a matter-wave interferometer based on Raman transitions which simultaneously interrogates two different atomic species ($^{87}$Rb and $^{85}$Rb). The simultaneous aspect of our experiment presents encouraging preliminary results for future dual-species atom interferometry projects and seems very promising by taking advantage of a differential acceleration measurement. Indeed the resolution of our differential accelerometer remains lower than $3.9\times 10^{-8}g$ even with vibration levels up to $1\times10^{-3}g$ thanks to common-mode vibration noise rejection . An atom based test of the Weak Equivalence Principle has also been carried out leading to a differential free fall measurement between both isotopes of $\Delta g/g = (1.2 \pm 3.2)\times10^{-7}$.
\end{abstract}


\maketitle




Light pulse atom interferometers \cite{Borde1989,Berman1997} have proven to be very high performance sensors with the development in the last decades of cold atom gravimeters \cite{Peters2001}, gravity gradiometers \cite{McGuirk2002} and gyroscopes \cite{Gustavson1997}. In addition to the undeniable contribution they could bring in practical applications such as inertial navigation and geophysics, they appear very promising for exploring fundamental physics such as for the determination of the fine structure constant \cite{Bouchendira2011}, the gravitationnal constant \cite{Fixler2007,Lamporesi2008}, but also for testing the Einstein's theory of general relativity with quantum objects \cite{Dimopoulos2007}. In that field, atom interferometers seem notably promising for detecting gravitational waves \cite{Dimopoulos2008}, exploring short range forces \cite{Ferrari2006,Wolf2007} and testing the Weak Equivalence Principle (WEP) \cite{Fray2004}.

In the context of testing the WEP, some projects under development aim to measure the acceleration of two different atomic species during few seconds of free fall in order to achieve highly sensitive measurements as it can be obtained in 10 m tall atomic fountains \cite{Dimopoulos2007}, drop towers \cite{Muntinga2013}, sounding rockets, parabolic flights \cite{Geiger2011} and satellites \cite{STE-QUEST}. To date, a single atom based ground test of the WEP was carried out by alternatively handling both isotopes of rubidium \cite{Fray2004}. This method, providing a non simultaneous differential measurement, exhibit a sensitivity limited by vibration noise, such as state of the art gravimeters \cite{Muller2008,LeGouet2008}. A special interest must thus be paid to develop atom interferometers which will simultaneously interrogate two different atomic species in order to take full advantages of a differential measurement and to achieve the targeted sensitivity and accuracy.

In this letter, we report the realization and the associated measurements of a dual-species atom interferometer giving access to a simultaneous measurement of accelerations undergone by both stable isotopes of rubidium. This experiment paves the way of future ground and space experiments dedicated to test the WEP.


For simultaneously measuring the acceleration undergone by $^{87}$Rb and $^{85}$Rb atoms, we use a Mach-Zehnder type atom interferometer consisting in a sequence of three equally spaced light pulses driving stimulated Raman transitions between the two fundamental hyperfine states of the atoms \cite{Kasevich1992}. The phase difference between both paths of the interferometer associated to the isotope $i$ is given to first approximation by:
\begin{equation}\label{interferometer phase}
\Delta \phi_{i} = (k_{\scriptsize\textrm{eff}}^{i}g_{i} - 2\pi \alpha)T^{2}
\end{equation}
where $k_{\scriptsize\textrm{eff}}^{i}$ is the effective wave vector of the Raman transition for isotope $i$, $g_{i}$ is the acceleration of gravity undergone by isotope $i$, $\alpha$ is the microwave chirp applied on the Raman frequency to compensate Doppler effect and $T$ is the time between two pulses of light.

The experimental setup is mainly derived from \cite{Bidel2013}. Atoms fall over a distance of about 6 cm and the Raman laser beam is retro-reflected by a mirror representing here our mutual inertial reference for both isotopes. The whole sensor is mounted on a passive vibration isolation table (Minus-K).

The cooling stage and the interferometric sequence are performed by a frequency doubled Telecom laser system \cite{Carraz2009}. All components involved at $1560$ nm come from fibered Telecom technology allowing the overall laser system to be compact and robust. The laser lines for the dual-species experiment are synthesized by phase modulation at $1560$ nm. During the cooling stage, the carrier frequency is tuned on the $^{87}$Rb cooling transition. The $^{85}$Rb cooling line and both repumper lines are generated by three modulation frequencies injected into the phase modulator as following:
f$_{\scriptsize\textrm{cooling}}^{~85}$ = f$_{\scriptsize\textrm{carrier}}$ + 1.126 GHz,
f$_{\scriptsize\textrm{repumper}}^{~87}$ = f$_{\scriptsize\textrm{carrier}}$ + 6.568 GHz,
f$_{\scriptsize\textrm{repumper}}^{~85}$ = f$_{\scriptsize\textrm{carrier}}$ + 1.126 + 2.915 GHz.
Approximatively one third of the global power is contained in additionnal modulation lines far from any atomic resonances. Phase modulation is also used for generating the laser lines during the interferometric sequence. Both Raman pairs are generated by directly injecting the Raman difference frequencies associated to both isotopes (\textit{i.e.} 6.834 GHz for $^{87}$Rb and 3.035 GHz for $^{85}$Rb), making the carrier frequency common to both Raman pairs. The Raman pair corresponding to $^{87}$Rb is red detuned by 0.59 GHz with respect to the excited hyperfine state $F'=2$ and therefore the one corresponding to $^{85}$Rb is red detuned by 1.86 GHz with respect to $F'=3$.

With this setup, approximatively $6\times10^{8}$, respectively $8\times10^{8}$, atoms of $^{87}$Rb, resp. $^{85}$Rb, are simultaneously loaded from a background vapor into a 3D Magneto-Optical Trap (MOT) in 250 ms. The atoms are then further cooled down in an optical molasses phase of 28 ms. At this point, a temperature of 1.3 $\mu$K for $^{87}$Rb and 2.1 $\mu$K for $^{85}$Rb is measured by Raman spectroscopy. Additionnal trap loss collisions due to interspecies atomic collisions \cite{Suptitz1994} do not exceed 10-15\% in our case. These results show that the additional laser lines do not have any significant impacts on the cooling efficiency. Atoms are then prepared in the Zeeman sublevel $m_{F}=0$ of the hyperfine ground state ($F=1$ for $^{87}$Rb or $F=2$ for $^{85}$Rb) thanks to a microwave pulse selection.

During the free-fall, the interferometric sequence occurs in a vertical uniform magnetic field of 28 mG. The sequence consists in three Raman laser pulses of durations 2 - 4 - 2 $\mu$s equally spaced appart in time by $T=40$ ms. The Raman laser pulses couple at the same time the states $|F\!=\!1,m_{F}\!=\!0\!>$ to $|F\!=\!2,m_{F}\!=\!0\!>$ for $^{87}$Rb and $|F\!=\!2,m_{F}\!=\!0\!>$ to $|F\!=\!3,m_{F}\!=\!0\!>$ for $^{85}$Rb. The same microwave chirp $|\alpha| \simeq 25.143$ MHz is applied to both Raman difference frequencies in order to compensate the time-dependant Doppler shift induced by gravity.

Finally, the atomic population repartition between the two coupled states is measured for each species by fluorescence detection. The atomic cloud is illuminated by two successive sequences of three light pulses of durations 1.5 - 0.05 - 1.5 ms. The first sequence induces the fluorescence signal from $^{87}$Rb atoms: the first pulse detects atoms in $F=2$, the second one fully transfers atoms from $F=1$ to $F=2$, the third pulse is identical to the first one and detects atoms intially in $F=1$. Following the first one, a second equivalent sequence is realized for $^{85}$Rb.

The whole sequence is performed at a repetition rate of 2.5 Hz.

The two signals from the dual-species atom interferometer are sinusoid functions of the interferometric phase and can be expressed as:
\begin{equation}\label{parametric ellipse}
\left\{
	\begin{array}{ll}
			\vspace{0cm}
			\frac{P_{87}-P_{87}^{0}}{C_{87}} = \cos(\Delta\phi_{87})
			\\
			\frac{P_{85}-P_{85}^{0}}{C_{85}} = \cos(\Delta\phi_{87}+\phi_{d})
	\end{array}
\right.\
\end{equation}
where $P_{87{,}85}$ are the proportions of atoms in the upper hyperfine ground state, $P^{0}_{87{,}85}$ are the offsets of the population measurements, $C_{87{,}85}$ are the fringe visibilities, $\Delta\phi_{87}$ is the interferometric phase for $^{87}$Rb as expressed in Eq.(\ref{interferometer phase}) and $\phi_{d}$ represents the differential phase between the two species. The interferometric phases can also be expressed without approximation using the atom interferometer response functions $f_{87,85}$ \cite{Geiger2011} associated to each isotope:
\begin{equation}\label{response functions}
\left\{
	\begin{array}{l}
	\vspace{0.1cm}
			\Delta\phi_{87}=\int f_{87}(t)(k_{\scriptsize\textrm{eff}}^{87}a_{87}(t) - 2\pi \alpha)\,dt+\phi^{87}_{SE}
			\\
			\Delta\phi_{85}=\int f_{85}(t)(k_{\scriptsize\textrm{eff}}^{85}a_{85}(t) - 2\pi \alpha)\,dt+\phi^{85}_{SE}
			\\
			~~~~~~~=\Delta\phi_{87}+\phi_{d}
	\end{array}
\right.\
\end{equation}
with $a_{87}(t)=g_{87}-\tilde{a}(t)$ and $a_{85}(t)=g_{87}+\Delta g-\tilde{a}(t)$. $\phi^{87,85}_{SE}$ are the phase shifts due to systematic effects, $\tilde{a}$ represents the vibrations of the Raman mirror and $\Delta g$ is the WEP violation signal.   

In order to compare the free fall of both isotopes we need to focus on the expression of $\phi_{d}$ which is the sum of all differential terms listed in Table \ref{differential phase}. Term 1 is the WEP violation signal impacted by the measurement scale factor. Term 2 corresponds to the difference of scale factor coming from the wave vector difference $\delta k=k_{\scriptsize\textrm{eff}}^{85}-k_{\scriptsize\textrm{eff}}^{87}$. Terms 3 and 4 limit common-mode vibration noise rejection coming respectively from the difference in response functions and from $\delta k$. Moreover these two terms do not contribute significantly to the accuracy of $\phi_{d}$ determination. Term 5 contributes to the phase shift when $\alpha$ does not compensate exactly the Doppler shift induced by gravity. This contribution is negligible in our experiment. Term 6 gathers all the phase shifts due to other differential systematic effects (see Table \ref{systematic effects}).

\begin{table}
\caption
{
\label{differential phase}
Phase shift terms composing the expression of the differential phase $\phi_{d}$. The first column presents phase shift terms according to the response function viewpoint. The second column presents the dominant term of the expansion of the previous terms in the low frequency limit ($\omega \rightarrow 0$). $f_{87,85}$ are triangle like funtions verifying $\int f_{87},_{85}(t)\,dt \simeq T^{2}$, $\delta\!f(t)=f_{87}(t)-f_{85}(t)$ and $2\pi\alpha_{0}=k_{\scriptsize\textrm{eff}}^{87}g_{87}$.
}
\begin{ruledtabular}
\begin{tabular}{ccc}

			\rule[0cm]{0pt}{0.2cm} Term & Phase Shift & Dominant Term\\
			\rule[0cm]{0pt}{0.2cm}  &  &($\omega \rightarrow 0$)\\
			\hline
			
			\rule[0.2cm]{0pt}{0.2cm} 1 & $(k_{\scriptsize\textrm{eff}}^{87}-\delta k)\Delta g \int f_{85}(t)\,dt$ & $(k_{\scriptsize\textrm{eff}}^{87}-\delta k) \Delta g T^{2}$\\

			\rule[0.2cm]{0pt}{0.2cm} 2 & $\delta k g_{87} \int f_{85}(t)\,dt$ & $\delta k g_{87}T^{2}$\\
			
			\rule[0.2cm]{0pt}{0.2cm} 3 & $k_{\scriptsize\textrm{eff}}^{87}\int \delta\!f(t) \tilde{a}(t)\,dt$ & $k_{\scriptsize\textrm{eff}}^{87}\tilde{a}T(\frac{2}{\Omega_{87}}-\frac{2}{\Omega_{85}})$\\

			\rule[0.2cm]{0pt}{0.2cm} 4 & $-\delta k\int f_{85}(t)\tilde{a}(t)\,dt$  &  $-\delta k \tilde{a}T^{2}$\\
			
			\rule[0.2cm]{0pt}{0.2cm} 5 & $2 \pi(\alpha\!-\!\alpha_{0})\!\int \delta\!f(t)\,dt$ & $2 \pi(\alpha\!-\!\alpha_{0})T(\frac{2}{\Omega_{87}}-\frac{2}{\Omega_{85}})$\\ 
			
			\rule[0.2cm]{0pt}{0.2cm} 6 & $\phi^{85}_{SE}-\phi^{87}_{SE}$ &  $\phi^{85}_{SE}-\phi^{87}_{SE}$\\

\end{tabular}
\end{ruledtabular}
\end{table}

\begin{figure*}
\centerline{\includegraphics[width=14cm]{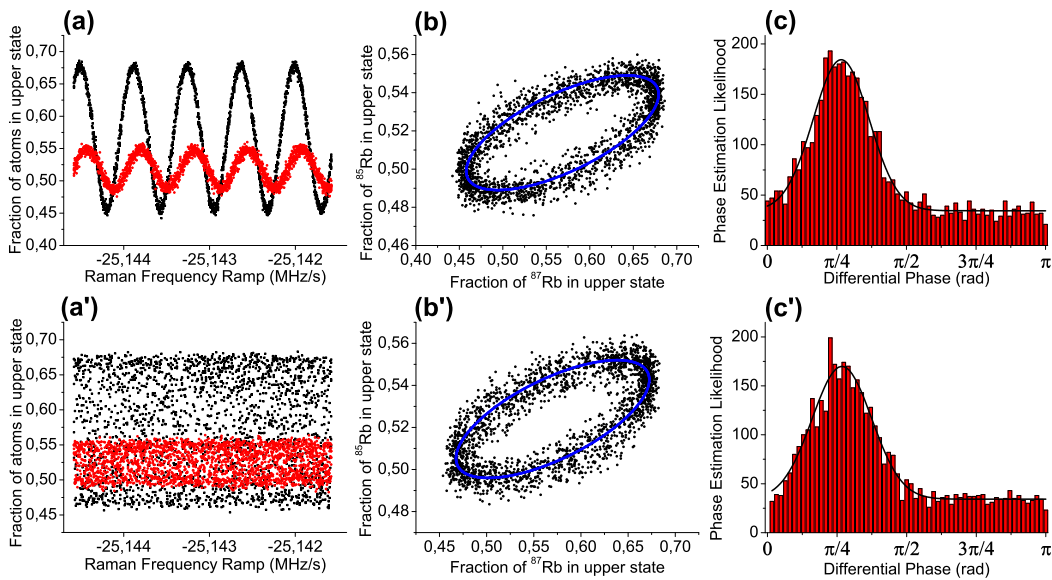}}
\caption
{
Typical simultaneous dual-species interferometric signals. On the top row, signals from a low vibration noise environment ($\sigma_{\tilde{a}}^{rms}=2.4\times10^{-7}g$), on the bottom row, signals from a high vibration noise environment ($\sigma_{\tilde{a}}^{rms}=1\times10^{-3}g$) in the atom interferometer bandwidth (0$\rightarrow$12.5 Hz). (a) and (a'): interference signals for $T=40$ ms as a function of the microwave chirp $\alpha$, in black dots from $^{87}$Rb, in red dots from $^{85}$Rb. (b) and (b'): interference signals from $^{85}$Rb are plotted versus interference signals from $^{87}$Rb drawing the ellipses fitted by the solid blue line curves. (c) and (c'): $\phi_{d}$ is derived from a gaussian fit, solid black line, of the phase estimation likelihood (red histogram) estimated by ``Direct Phase Extraction'' method. Each of the $N=2200$ data points represents a single drop of the atoms at a repetition rate of $2.5$ Hz. The resolution on $\phi_{d}$ is $8$ mrad in case 1, corresponding to $3.2\times10^{-8}g$ and $10$ mrad in case 2, corresponding to $3.9\times10^{-8}g$. This leads to a common-mode noise rejection ratio of $20\log(\frac{1\times10^{-3}}{\sqrt{N}\times3.9\times10^{-8}})=55$ dB corresponding to a rejection factor of 550.
}
\label{interferometric signals}
\end{figure*}

Typical simultaneous dual-species fringes are reported in Fig.\ref{interferometric signals}.a. A non zero differential phase $\phi_{d}$ is clearly observed at first sight between fringes of $^{87}$Rb and $^{85}$Rb, mainly coming from terms 2 and 6. Moreover, the fringe visibility is lower for $^{85}$Rb which is due to the generation of Raman laser lines by modulation \cite{Carraz2012}. Indeed the additional laser lines induce destructive interferences of the transition probability depending on the distance between the atoms and the Raman mirror during light pulses. This effect is more significant for the $^{85}$Rb because of its narrower hyperfine structure thus degrading its associated signal visibility.


Table \ref{systematic effects} presents the main contributions impacting $\phi_{d}$ determination. $\phi_{d}$ is here extracted from a sinus fitting of fringes after averaging signals over 15 min and for the two reversed directions of $\overrightarrow{k_{\scriptsize\textrm{eff}}}$.

\begin{table}
\caption
{
\label{systematic effects}
Main contributions affecting the differential acceleration measurement.
}
\begin{ruledtabular}
\begin{tabular}{l d d}

		 	\rule[0cm]{0pt}{0.2cm} & \Delta g/g & \rm{Uncertainty}\\
			&\times10^{-7}&\times10^{-7}\\
			\hline
			\hline

			\rule[0.05cm]{0pt}{0.3cm} Exp. Results & -27.6 & 0.25\\
			\hline
			
			\rule[0.1cm]{0pt}{0.2cm} Term 2 ($\delta k/k_{\scriptsize\textrm{eff}}$)& 49.4 & 0\\

			\rule[0cm]{0pt}{0.6cm} Term 6 Correction:& &\\

			\rule[0cm]{0pt}{0.2cm} ~~~Additionnal Lines& -23.3 & 1.1\\

			\rule[0cm]{0pt}{0.2cm} ~~~Frequency Shifts& 0.3 & 2.9\\

			\rule[0cm]{0pt}{0.2cm} ~~~Coriolis Effect& 0 & 0.6\\

			\rule[0cm]{0pt}{0.2cm} ~~~WaveFront Aberrations& 0 & 0.1\\
			\hline

			\rule[0.2cm]{0pt}{0.2cm} Total & 1.2 & 3.2\\

\end{tabular}
\end{ruledtabular}
\end{table}

The systematic effect due to additionnal laser lines, coming from the use of a phase modulated Raman laser, is corrected as in \cite{Carraz2012}. The uncertainty mainly comes from the uncertainties in the initial distance between the atomic clouds and the retro-reflection mirror (estimated at $\pm0.2$ mm) and in the difference of initial vertical velocities of both atomic species when the MOT is turned off (estimated $\leq$ 6 mm/s by Raman spectroscopy).

AC Stark shifts and second order Zeeman shift generate a phase shift by modifying the Raman difference frequencies. One-photon light shift and second order Zeeman shift are strongly minimized thanks to the averaging over reversed directions of $\overrightarrow{k_{\scriptsize\textrm{eff}}}$. Nevertheless the two-photon light shift \cite{Landragin2008} still affects the phase of the atom interferometer and the predicted error of this effect is calculated. All these frequency shifts are treated together ($cf.$ Table \ref{systematic effects}). The variation of the atomic transition frequency is measured by Raman spectroscopy during different measurement sessions and for reversed directions of $\overrightarrow{k_{\scriptsize\textrm{eff}}}$. The uncertainty of $2.9\times10^{-7}$, reported in Table \ref{systematic effects}, corresponds to the upper bound frequency variation over all these measurements and is mainly caused by Raman laser power fluctuations.

The transverse velocity of atoms makes the interferometer sensitive to Earth rotation through Coriolis effect. The differential transverse velocity is assumed to vary as the vertical one: the assumption of a random variation of $\pm6$ mm/s leads to an  uncertainty of $6\times10^{-8}$.

At the end of the cooling stage, $^{87}$Rb and $^{85}$Rb atoms exhibit a difference in temperature of approximatively 1 $\mu$K. Both isotopes will be then affected in a different way by wavefront aberrations of the Raman laser beam. The measured wavefront curvature creates an uncertainty of one order of magnitude lower than the other systematic effects \cite{Bidel2013}.

All these contributions lead to a final relative differential acceleration of $\Delta g/g = (1.2 \pm 3.2)\times10^{-7}$ with a resolution of $2.5\times 10^{-8}$.


This differential acceleration measurement, where both species simultaneously fall in a common reference frame, allows us to benefit from an efficient common-mode noise rejection. This kind of rejection, in an atom sensor, has only been reported to our knowledge in a gravity gradiometer handling two cold atomic clouds of Cs separated by a distance of about 1 m \cite{McGuirk2002}. We study here more specifically vibration rejection noise ($cf.$ terms 3 and 4) with two embedded clouds of $^{87}$Rb and $^{85}$Rb simultaneously submitted to the same light pulse interferometric sequence.

The two signals from the dual-species atom interferometer are sinusoid functions and thus parametrically describe an ellipse ($cf.$ Eq.(\ref{parametric ellipse})). Fig.\ref{interferometric signals}.b shows the ellipse obtained by plotting the interferometric signal from $^{85}$Rb versus the one from $^{87}$Rb. This highlights the correlation between those two signals: common-mode vibration noise and $\alpha$ chirp distribute the data points around the ellipse.

The sensitivity of the dual atom interferometer to vibrations has been tested by shaking the instrument platform on which the retro-reflection mirror is mounted. When vibration noise is much lower than typically half of the fringe spacing, corresponding to $12~\mu g$, fringes remain observable ($cf.$ Fig.\ref{interferometric signals}.a), otherwise fringes are blurred ($cf.$ Fig.\ref{interferometric signals}.a'). Nevertheless, as this vibration noise is common to both isotopes, the ellipse is still clearly visible ($cf.$ Fig.\ref{interferometric signals}.b'). By fitting of this ellipse, the differential phase shift $\phi_{d}$, and thus the differential acceleration, can still be derived.

For the data processing we use an ellipse fitting method called ``Direct Phase Extraction'' \cite{Wu_thesis}. Indeed this method is quick and easy to implement and chiefly allows a bias-free differential phase extraction. After extracting population offsets and fringe visibilities, from Eq.(\ref{parametric ellipse}) the common interferometric phase $\Delta\phi_{87}$ can be eliminated and $\phi_{d}$ is ``directly'' derived. Phase noise is usually assumed to be normally distributed, thus the optimal differential phase $\phi_{d}$ is given by the maximum of the phase estimation likelihood function,  reported in Fig.\ref{interferometric signals}.c and c'. A gaussian fit of this function produces the same result for $\phi_{d}$ as the sinus fitting method. The resolution remains less than 10 mrad (corresponding to $3.9\times10^{-8}g$) over a vibration noise range from $2.4\times10^{-7}g$ to $1\times10^{-3}g$ in the atom interferometer bandwidth (0$\rightarrow$12.5 Hz). Over this range the resolution remains almost constant and seems limited by amplitude noises on the interferometric signals. For vibration levels one order of magnitude higher the ellipse begins to blur. These results exhibit a common-mode vibration noise rejection ratio higher than 55 dB corresponding to a rejection factor of 550.

Terms 3 and 4 of Table \ref{differential phase} are responsible for vibration noise rejection efficiency. Term 4 gives the maximum rejection ratio achievable corresponding to $k_{\scriptsize\textrm{eff}}/\delta k=2\times10^{5}$($\equiv106$ dB). However, the rejection is currently limited by term 3. When considering constant accelerations of the Raman mirror, this term reduces to first order in $1/\Omega_{i}$ to the expression shown in Table \ref{differential phase}, corresponding to the differential transfer function \cite{Cheinet2008} of the dual atom accelerometer in the low frequency limit. This term depends on the Rabi frequencies ($\Omega_{i}$) associated to the Raman transitions of both $^{87}$Rb and $^{85}$Rb interferometers: the vibration rejection ratio is thus directly determined by the mismatch of the Rabi frequencies. We measured a maximum mismatch of a factor 4 between $\Omega_{87}$ and $\Omega_{85}$, mainly due the additionnal laser lines generated by phase modulation, limiting the rejection to 78 dB. This estimation demonstrates that our measurement is limited by an additional source of noise ($e.g.$ detection noise) and not by the efficiently rejected vibration noise.


To conclude, we have demonstrated the realization of a simultaneous dual-species atom accelerometer based on light pulse atom interferometry. This experiment relies on the use of a compact and robust laser system based on a single laser diode for efficiently cooling and simultaneously manipulating both isotopes of rubidium. A new atom based test of the WEP has been carried out leading to a differential free fall measurement $\Delta g/g = (1.2 \pm 3.2)\times10^{-7}$. The simultaneous aspect of this experiment takes advantage of a differential measurement by allowing an efficient common-mode vibration noise rejection higher than 55 dB, the resolution of the differential accelerometer remaining less than $3.9\times 10^{-8}g$ even for vibration levels up to $1\times10^{-3}g$. Futhermore, the detrimental effects of additionnal laser lines on the dual interferometer could be greatly reduced by implementing this experiment in a microgravity environment where the position between the atomic clouds and the Raman mirror do not change during the free fall \cite{Carraz2012}. Finally, these results demonstrate the feasibility of a simultaneous dual-species atom accelerometer paving the way of future ground and space based experiments dedicated to test the WEP.

\begin{acknowledgments}
This work was supported by the French Defense Agency (DGA).
\end{acknowledgments}

\bibliography{bibliography_of_Simultaneous_Dual_Atom_Interferometry_3.0}

\end{document}